\documentclass[useAMS,usenatbib,a4paper]{mnras} 

\usepackage{graphicx}
\usepackage{amssymb}
\usepackage{subfigure}
\usepackage{captcont}
\usepackage{multirow}
\usepackage{longtable}
\usepackage{comment}
\usepackage{url}
\usepackage{lscape}
\usepackage{booktabs,multirow}
\usepackage{times} 
\usepackage{txfonts}
\usepackage{aas_macros} 

\usepackage[usenames]{color}

\newcommand{\cd}{d$^{-1}$}

\newcommand{\Msun}{M$_{\odot}$}
\newcommand{\Lsun}{L$_{\odot}$}

\title[Distorted roAp pulsation in J0855]{LCO observations of a super-critical distorted pulsation in the roAp star J0855 (TYC\,2488-1241-1)}
\author[D. L. Holdsworth et al.]{Daniel L. Holdsworth,$^{1,2}$\thanks{E-mail:dlholdsworth@uclan.ac.uk}
Hideyuki Saio,$^{3}$
Ramotholo R. Sefako$^{4}$ and
\newauthor Dominic M. Bowman${^5}$\\
$^{1}$ Jeremiah Horrocks Institute, University of Central Lancashire, Preston PR1 2HE, UK\\
$^{2}$ Department of Physics, North-West University, Mafikeng Campus, Private Bag X2046, Mmabatho 2745, South Africa\\
$^{3}$ Astronomical Institute, School of Science, Tohoku University, Sendai 980-8578, Japan\\
$^{4}$ South African Astronomical Observatory, PO Box 9, Observatory, Cape Town 7935, South Africa\\
$^{5}$ Instituut voor Sterrenkunde, KU Leuven, Celestijnenlaan 200D, 3001 Leuven, Belgium\\}

\begin{document}

\date{\today}

\pagerange{\pageref{firstpage}--\pageref{lastpage}} \pubyear{2018}

\maketitle

\label{firstpage}

\begin{abstract}
We report the results of a 60-hr photometric campaign of a rapidly oscillating Ap star, J0855 (TYC\,2488-1241-1). We have utilised the multi-site Las Cumbres Observatory's (LCO) 0.4-m telescopes to obtain short cadence $B-$band observations of an roAp star previously lacking detailed study. Our observations confirm the rotation period presented in the discovery paper of this star ($P_{\rm rot}=3.0918$\,d), and reveal the star to be pulsating in a distorted mode. The $B$ data show this star to be among the highest amplitude roAp stars, with a peak-to-peak amplitude of 24\,mmag. Modelling of the pulsation frequency at $197.2714$\,\cd\, (2283\,$\muup$Hz; $P=7.30$\,min) shows that this star belongs to the subgroup of super-critical pulsators, where the observed frequencies are above the theoretical acoustic cutoff frequency. From the modelling, we deduce that the star's rotation axis is inclination angle of about $30^\circ$ to the line-of-sight, with an angle of obliquity of the magnetic axis to the rotation axis of either $40^\circ$ or $24^\circ$ depending on whether the pulsation mode is dipole or quadrupole, respectively.
\end{abstract}


\section{Introduction}
\label{sec:intro}

The rapidly oscillating Ap (roAp) stars, which are found at the base of the classical instability strip, are a class of pulsating star with only a few dozen members. These stars, discovered by \citet{kurtz82}, show pulsational variability with periods between about five and 25\,min \citep{joshi16}. 

The Ap stars, as a class, show enhanced features of Sr, Cr and/or Eu in classification resolution spectra, with high resolution spectroscopy revealing overabundances of elements such as Pr, Nd, Gd and Tb up to a million times the solar value \citep{ryabchikova04}. It is thought that the presence of a strong, stable, global magnetic field freezes convection allowing for chemical stratification of these elements in the atmospheres of Ap stars. This chemical stratification forms spots or clouds in the stellar atmosphere which modulate the light output of the star over its rotation period, thus allowing the rotation period to be measured (assuming observations cover the rotation period). As a result of magnetic braking \citep{stepien00}, the Ap stars are more slowly rotating compared to their non-magnetic counterparts, showing rotation periods up to centuries \citep{mathys15}.

The pulsations in the roAp stars, of which 61 are discussed in the literature, are high-overtone p-modes, thought to be driven by the $\kappa$-mechanism acting in the H\,{\sc{i}} ionization zone \citep{balmforth01,saio05}. However, there are some roAp stars which pulsate with frequencies greater than their theoretical limit (the acoustic cutoff frequency). It has been suggested that these pulsations are not driven by the $\kappa$-mechanism, but rather by turbulent pressure \citep{cunha13}.

The pulsation geometry of the roAp stars is unique. Rather than having a pulsation axis aligned to the rotation one, it is more closely aligned to the magnetic one. This results in oblique pulsation which manifests itself in amplitude and phase variations in the measured pulsation mode over the rotation cycle of the star \citep[e.g. ][]{kurtz82,ss85a,ss85b, bigot02,bigot11}. Furthermore, in the case of a non-distorted mode the oblique pulsation results in a multiplet, in frequency space, of $2\ell+1$ peaks separated by the rotation frequency. Such an occurrence allows for the degree of the mode to be easily determined.

The subject of this paper, TYC 2488-1241-1 ($\alpha$: 08:55:22.21, $\delta$: $+32$:42:36.3 (J2000); $V=10.8$; hereafter J0855), was first reported as an roAp star by \citet{holdsworth14a} as a result of a survey of A stars in the SuperWASP archive. To date, three of the 11 roAp stars found in the SuperWASP archive \citep{holdsworth14a,holdsworth15} have been the subject of dedicated photometric campaigns \citep{holdsworth16,holdsworth18a,holdsworth18b}. Those observations found these stars to have distorted, high-amplitude quadrupole pulsations. It is not surprising that, when observed using $B$ filters, these stars have some of the largest amplitudes amongst the roAp stars: the broad filter employed by SuperWASP \citep[$4000-7000$\,\AA;][]{pollacco06} is not favourable to search for pulsations in Ap stars, therefore the WASP discoveries will be intrinsically high-amplitude pulsators so that their variability is detectable in the diluted photometry. These stars, therefore, have already expanded the parameter space which roAp pulsations are known to exist.

Here, we present new photometric observations of J0855 using telescopes of the Las Cumbres Observatory (LCO). We model the observations using the same method as employed for the other SuperWASP roAp stars studied in detail to compare J0855 to the family of roAp stars.

\section{Observations and reduction}

For the study of J0855, we have utilised the Las Cumbres Observatory's 0.4-m telescopes, with the SBIG STL-6303 cameras, located at Teide, McDonald and Haleakala observatories. Observing blocks were 2.6\,hr long to acquire about 20 pulsation cycles per visit. Each exposure was 30\,s and was acquired through a $B$ filter to maximise our signal-to-noise \citep{medupe98}.

The LCO provides science frames in either their raw format, or in a fully reduced state. For our analysis, we utilised their custom reductions which, in brief, consist of: bad pixel masking, bias subtraction, dark subtraction and flat field correction. The pipeline also performs aperture photometry, but we chose to perform our own extraction of sources. 

Aperture photometry was performed using the code described in \citet{holdsworth18b}: we used optimised elliptical apertures to extract the target and comparison star fluxes. Background measurements were made on a starless nearby, area on the CCD, using the target and comparison apertures, and subtracted from the stars, respectively.  Time stamps, in BJD (TDB), were created for each frame given the observatory location and frame pointing co-ordinates. 

In total, we obtained over 5050 useful data points over 35 different visits spanning nearly 86.29\,d. Due to weather and how LCO rank and schedule observations, not all blocks were fully completed. This is evident in some of our figures below, where frequencies and amplitudes are calculated over shorter data sets.

\section{Rotation analysis}
\label{sec:rot}

For the following analysis, we use differential magnitudes. Although the errors for differential observations are larger than the individual target/comparison data, the rotation period is known, and the rotation signature will dominate over the noise in the differential data.

The comparison star, TYC\,2488-1458-1 ($\alpha$: 08:55:32.33, $\delta$: $+32$:38:07.2 (J2000); $V=11.4$), is located approximately 300\,arcsec away from the target on the sky and is the closest suitable comparison star. Not much is known about this star, but photometric colours indicate it is likely an F or A star. Although the A and F stars lack a deep surface convection zone required to produce star spots, many (if not all) stars of this spectral type show low-frequency variability, often due to either g-mode pulsations or r-modes associated with rotation \citep[e.g.][]{murphy14,balona15,vanreeth16,saio18,bowman18}. We therefore analyse the light curve of the comparison star to search for such low-frequency signals. We find no signs of periodic variability in this star (at the precision of our data).

\begin{figure}
\includegraphics[width=\linewidth]{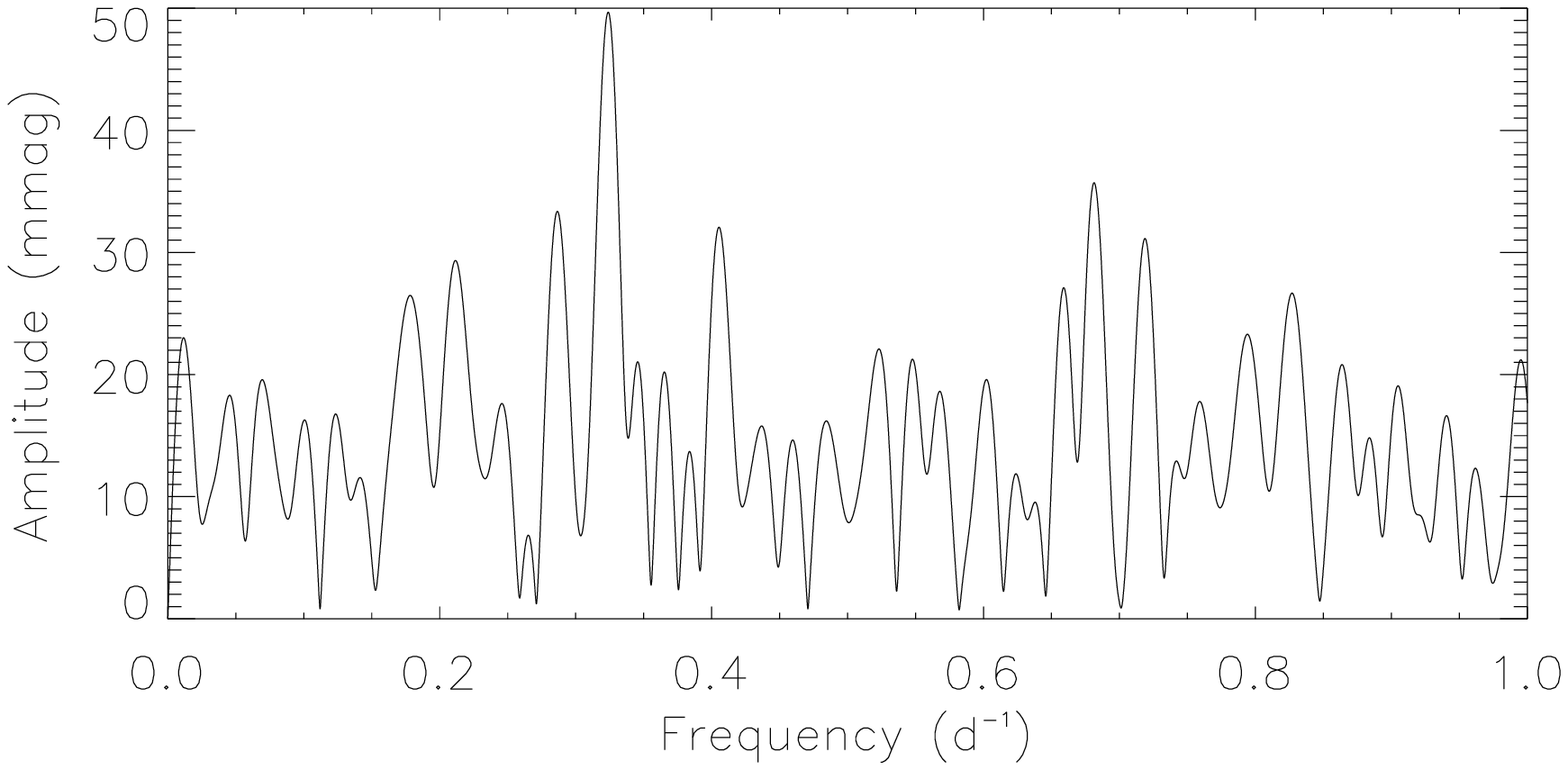}
\includegraphics[width=\linewidth]{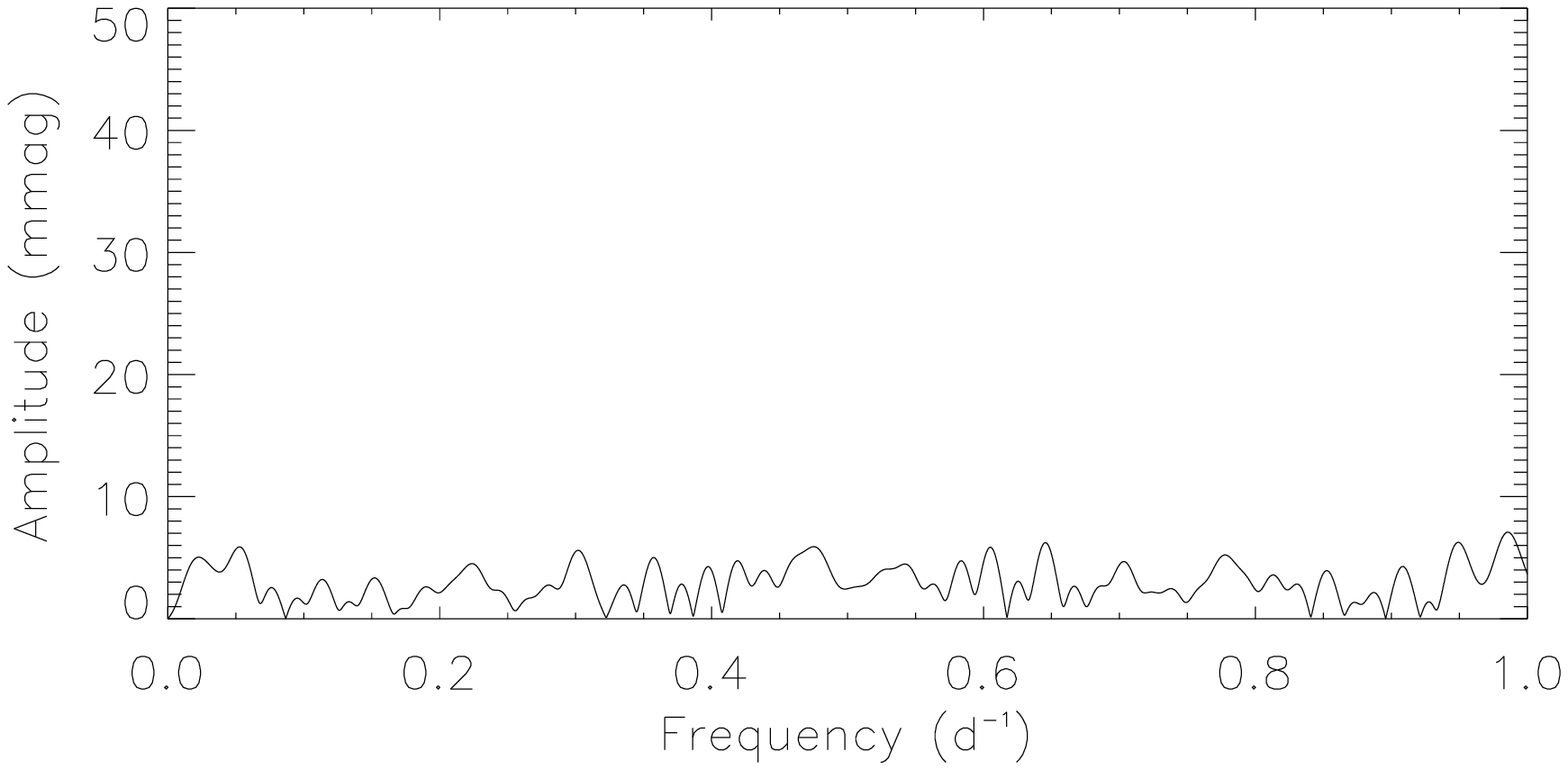}
\includegraphics[width=\linewidth]{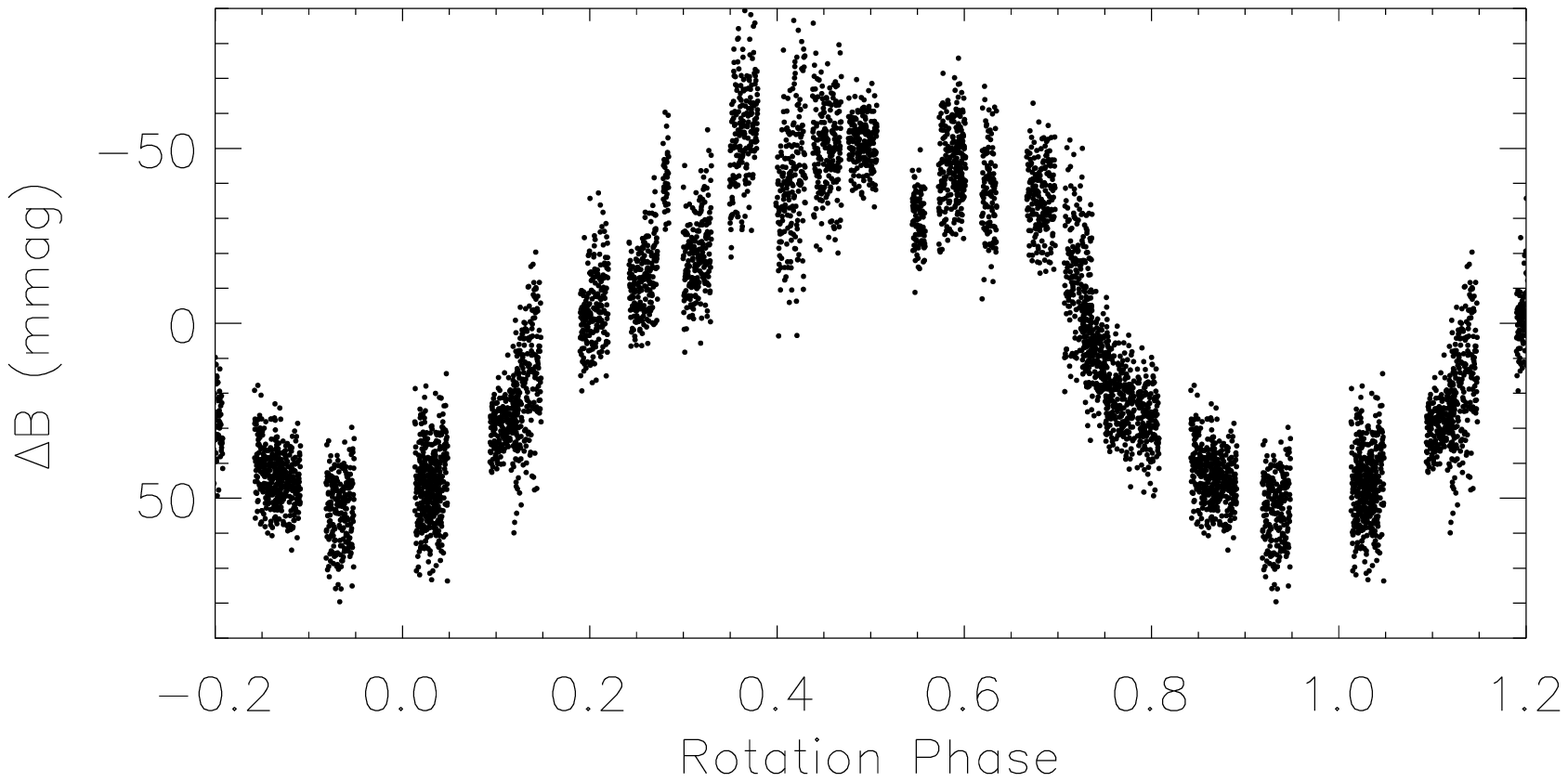}
\caption{Top: low frequency amplitude spectrum of J0855 showing the rotation frequency of the star at $0.32344\pm0.00017$\,\cd. Middle: amplitude spectrum of the residuals after pre-whitening the rotation signal from the data, showing the significance of the rotation peak in the top panel. Bottom: light curve folded on the rotation period of the star.}
\label{fig:rot}
\end{figure}

To calculate the rotation period of the star, we remove outlying points from the data, and calculate an amplitude spectrum to $1$\,\cd\, (Fig.\,\ref{fig:rot}, top). There is a strong signal at a frequency of about $0.32$\,\cd. We fit this frequency with non-linear least-squares to find the best fitting frequency to be $0.32344\pm0.00006$\,\cd, corresponding to a rotation period of $3.0918\pm0.0005$\,d. After pre-whitening this frequency from the data, we see there is no significant frequency left in the residuals; this is shown in the middle panel of Fig.\,\ref{fig:rot}. However, the error quoted on the frequency is the formal error calculated following \citet{montgomery99}. Realistic estimates of the frequency error can be found by comparing the amplitude error at low frequency to that at high frequency (as the frequency error is proportional to the amplitude error), where the noise is white. In doing so, we estimate the frequency error to be $3.1$ times greater than the formal error. Therefore, we determine the rotation period to be $3.0918\pm0.0016$\,d. Our results here are in agreement with those presented by \citet{holdsworth14a} who used SuperWASP photometry to determine the rotation period of J0855. We present a phase folded light curve in the bottom panel of Fig.\,\ref{fig:rot}. From Fig.\,\ref{fig:rot} it is evident that the rotation signature is not described by a single sinusoid which can be a trait of the Ap stars. There also appears to be a slight decrease in brightness around rotation phase 0.5. However, the precision of our data does not allow us to confidently make conclusions about this feature, although it could be a glimpse of the other pole's spot. Observations by the Transiting Exoplanet Survey Satellite \citep[{\it{TESS}}; ][]{ricker15} will provide more precise observations, covering several rotation periods, to investigate this feature.

From the rotational light variations, we are able to gain our first insight into the geometry of J0855. From the bottom panel of Fig.\,\ref{fig:rot} we are able to say that the sum of  $i$, the inclination angle, and $\beta$, the angle of obliquity between the rotation and the pulsation axes is $\lesssim90^{\circ}$ because, under the assumption that spots form at the poles of Ap stars (which is not always the case), we would otherwise see a doubly modulated rotational light variation. Although, as stated above, the dip in the light curve at about $\nu_{\rm rot}\sim0.5$ could be part of a spot at the other pole. This conclusion will be tested later in Section\,\ref{sec:modelling}. 

\section{Pulsation Analysis}
\label{sec:puls}

To analyse the pulsation in J0855, we use the non-differential photometry. Although losing data to poor conditions, this allows us to work with higher precision data as we are not including photometric error from the fainter comparison star.

\begin{figure}
\includegraphics[width=\linewidth]{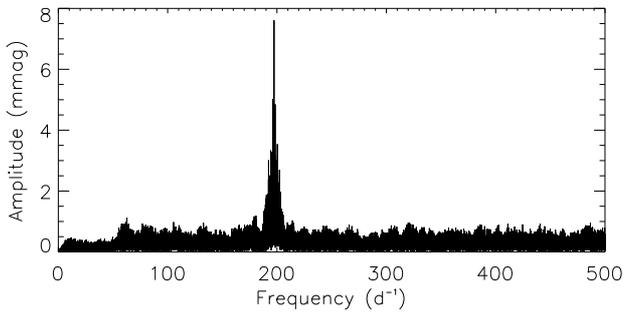}
\caption{Amplitude spectrum of the raw light curve of J0855, with low frequencies removed. At this scale, the structure around the pulsation are daily aliases.}
\label{fig:ft_all}
\end{figure}

We pre-whiten the data from each observing block up to a frequency of 50\,\cd\, and an amplitude of 2\,mmag. This procedure removes the effects of stellar rotation, sky transparency variations and low frequency noise from the light curve. The variability that we wish to study is suitably removed in frequency not to be affected by this process. We show an amplitude spectrum of all the data in Fig.\,\ref{fig:ft_all}. Clearly obvious is the pulsation at about $200$\,\cd.

\begin{figure}
\includegraphics[width=\linewidth]{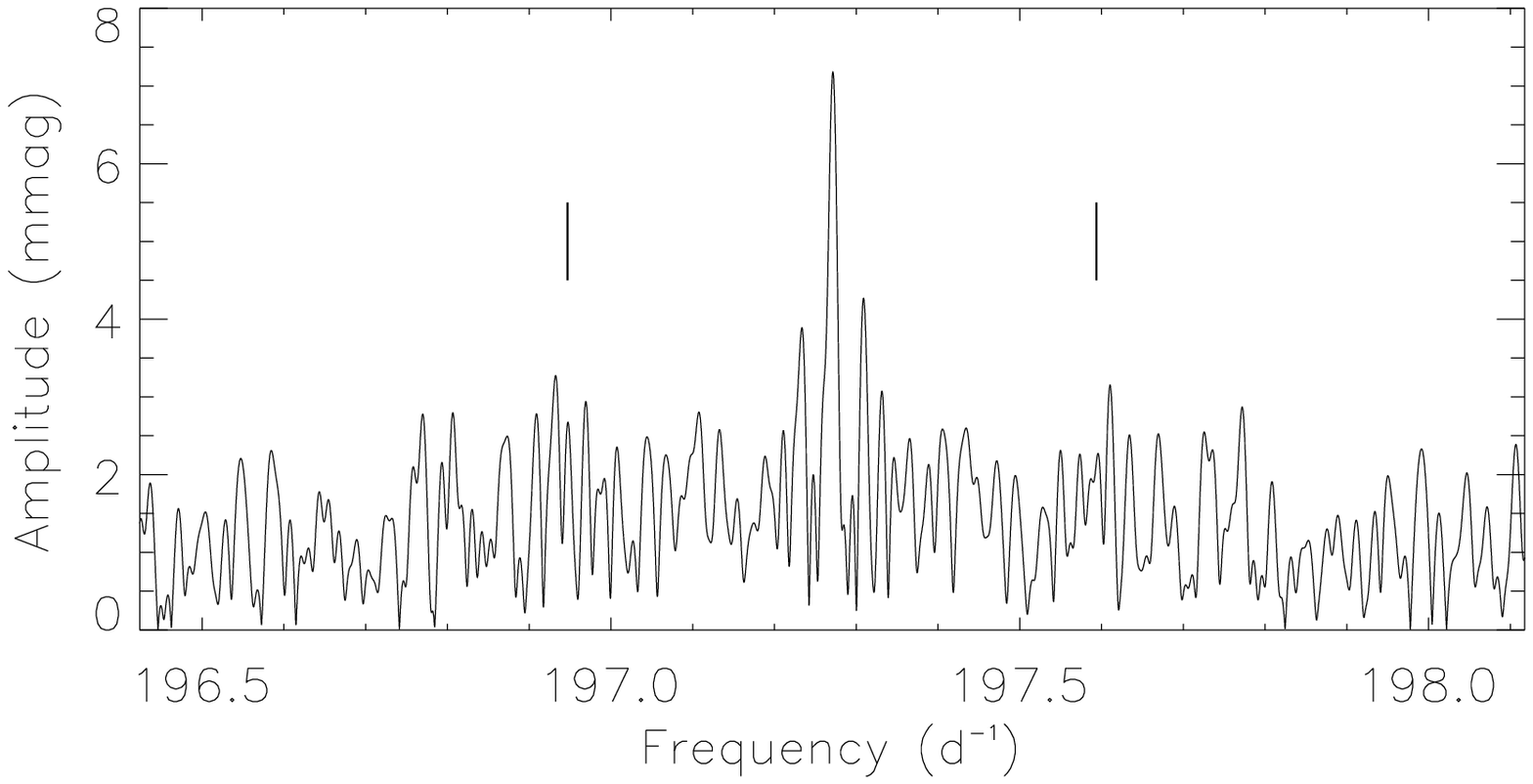}
\includegraphics[width=\linewidth]{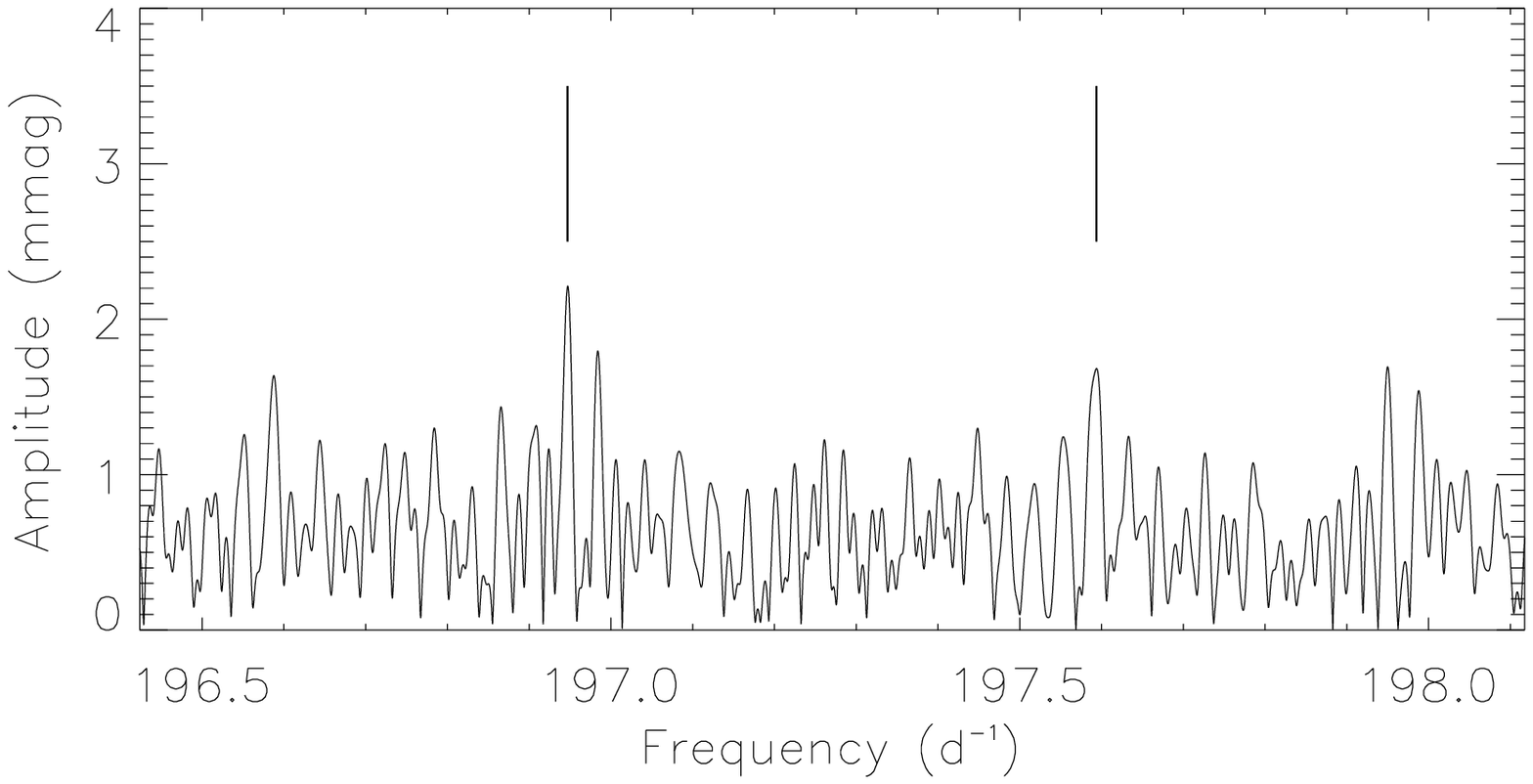}
\includegraphics[width=\linewidth]{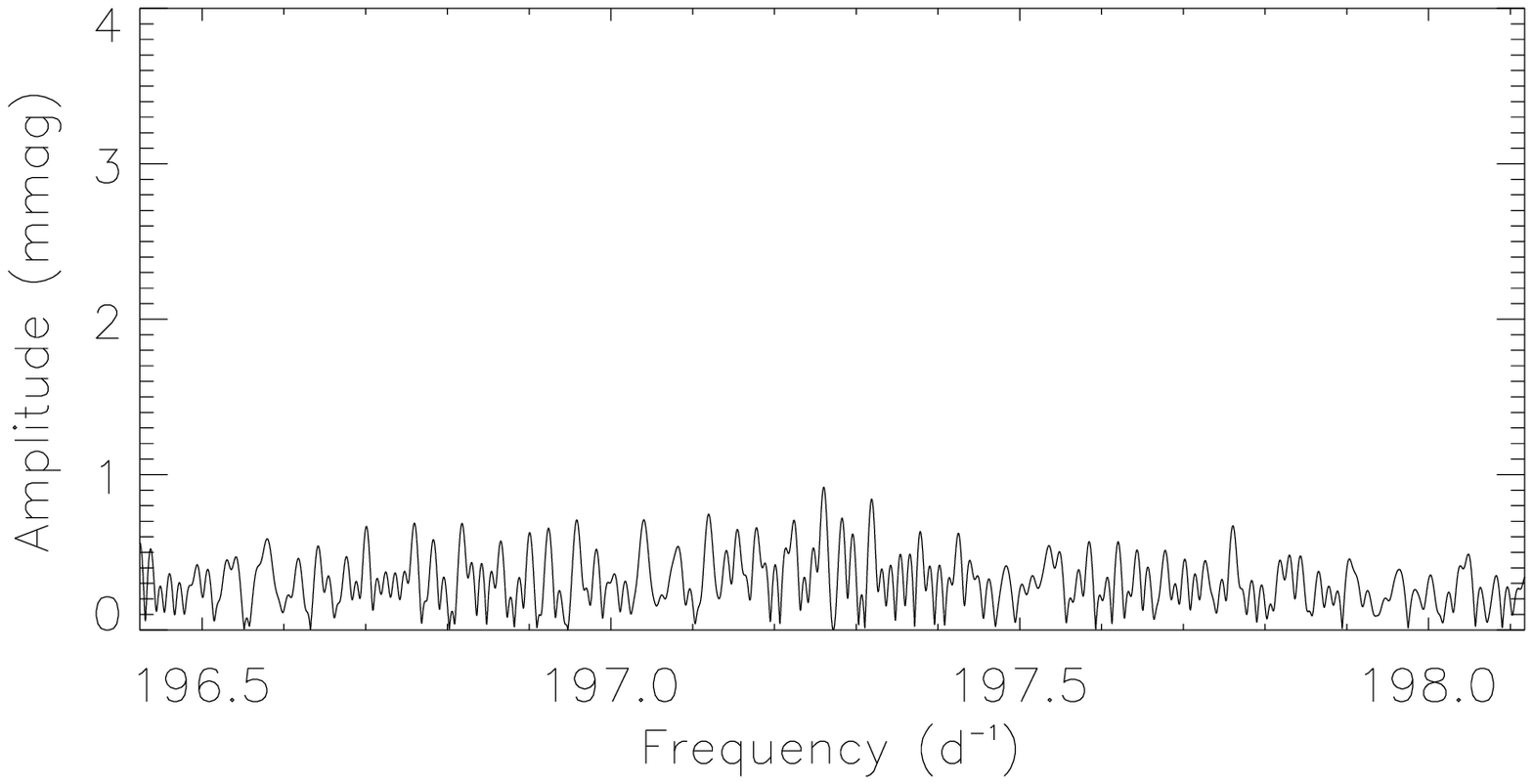}
\caption{Zoomed view of the amplitude spectra of the pulsation signature in J0855, and the subsequent steps of identifying and removing the rotational side lobes. Top: the pulsation frequency. Middle: identification of the rotational side lobes. The peaks around 196.6\,\cd\, and 198.0\,\cd\, are daily aliases of the side lobes. Bottom: amplitude spectrum of the residuals showing no significant peaks remain. The vertical bars in the top and middle panel show the expected frequencies of side lobes split by the rotation frequency. Note the change in the ordinate scale between the top and subsequent panels.}
\label{fig:pre-white-steps}
\end{figure}

After fitting the pulsation frequency with non-linear least-squares and pre-whitening it from the data, we identify two further peaks either side of the main pulsation frequency. These two side lobes are a result of the oblique pulsation seen in roAp stars, and should be split from the principal frequency by the rotation frequency of the star \citep{kurtz82}. We show the pulsation peak, and the subsequent process of fitting and removing the side lobes, in Fig.\,\ref{fig:pre-white-steps}.

The results of a non-linear least-squares fit of the frequencies to the data are shown in Table\,\ref{tab:freqs}. Beyond these three significant peaks, we find no other significant signs of variability in J0855, nor do we detect harmonics of the principal frequency. The highest noise peaks in the amplitude spectrum where the first harmonic would appear have amplitudes of about 0.5\,mmag, which may be limiting our ability to detect any harmonic.

\begin{table}
\caption{The non-linear least-squares fit results of the three frequencies to the light curve. The phases are calculated with a zero-point of BJD$-245000.0000=8226.6465$.}
\label{tab:freqs}
\begin{tabular}{lrrrr}
\hline
ID & \multicolumn{1}{c}{Frequency} & \multicolumn{1}{c}{Amplitude} & \multicolumn{1}{c}{Phase} \\
 & \multicolumn{1}{c}{(\cd)} & \multicolumn{1}{c}{(mmag)} & \multicolumn{1}{c}{(rad)} \\
\hline
$\nu - \nu_{\rm rot}$ 	&$196.9474\pm0.0005$ & $2.176\pm0.170$ & $-2.090\pm0.087$\\
$\nu $ 			& $197.2717\pm0.0002$ & $6.948\pm0.171$ & $-2.810\pm0.027$\\
$\nu + \nu_{\rm rot}$ &$197.5930\pm0.0007$ & $1.633\pm0.170$ & $2.897\pm0.115$\\

\hline
\end{tabular}
\end{table}

We note that, using the rotation frequency derived in Section\,\ref{sec:rot}, the sidelobes presented in Table\,\ref{tab:freqs} are not exactly split by the expected frequency, but are in agreement to 1.5 and 2.9\,$\sigma$ for the negative and positive lobes, respectively. Close inspection of the $\nu + \nu_{\rm rot}$ peak in the amplitude spectrum shows that it is not well represented by window function and is distorted, probably by noise. 

The oblique pulsator model predicts the presence of a triplet for a non-distorted dipole pulsation, and a quintuplet for a non-distorted quadrupole pulsation (i.e. a multiplet of $2\ell+1$). In the case of J0855, we observe a triplet in the amplitude spectrum and thus {\it{assume}} that J0855 pulsates in a dipole mode. 

Furthermore, the oblique pulsator model predicts that the side lobes are split by {\it{exactly}} the rotation frequency of the star, and that the phases of all components of the triplet should be equal. We therefore test this model by using the rotation frequency to forcibly split the side lobes. We then choose a zero-point in time where the phases of the side lobes will be equal and fit the triplet by linear least-squares. The results of this procedure, shown in Table\,\ref{tab:force_sidelobes}, reveal that J0855 is pulsating in a slightly distorted mode as, although close, $\phi_{-1}=\phi_{+1}\neq\phi_0$. We also note here that the amplitudes of the side lobes are unequal due to the Coriolis force \citep[e.g.][]{bigot02}.

\begin{table}
\caption{Results of fitting the triplet with frequencies separated by exactly the rotation frequency. The phases are calculated with a zero-point of BJD$-2450000.0000=8226.9166$.}
\label{tab:force_sidelobes}
\begin{tabular}{lrrrr}
\hline
ID & \multicolumn{1}{c}{Frequency} & \multicolumn{1}{c}{Amplitude} & \multicolumn{1}{c}{Phase} \\
 & \multicolumn{1}{c}{(\cd)} & \multicolumn{1}{c}{(mmag)} & \multicolumn{1}{c}{(rad)} \\
\hline
$\nu - \nu_{\rm rot}$	& $196.9483$ & $2.131\pm0.170$ & $	-0.737\pm0.080$\\
$\nu $			& $197.2717$ & $6.911\pm0.171$ & $	-0.961\pm0.025$\\
$\nu + \nu_{\rm rot}$	& $197.5951$ & $1.616\pm0.170$ & $	-0.737\pm0.105$\\
 \hline
\end{tabular}
\end{table}

The final test we apply to the data is to check for amplitude and phase variability of the pulsation. It is expected that the amplitude of the pulsation mode is modulated with the rotation period, which gives rise to the side lobes to the pulsation. We expect to see phase variations in the pulsation if a pulsation node crosses the line-of-sight, which in the case of a pure dipole mode is the equator and co-latitudes of $\pm54.7^\circ$ for a pure quadrupole mode. 

\begin{figure}
\includegraphics[width=\linewidth]{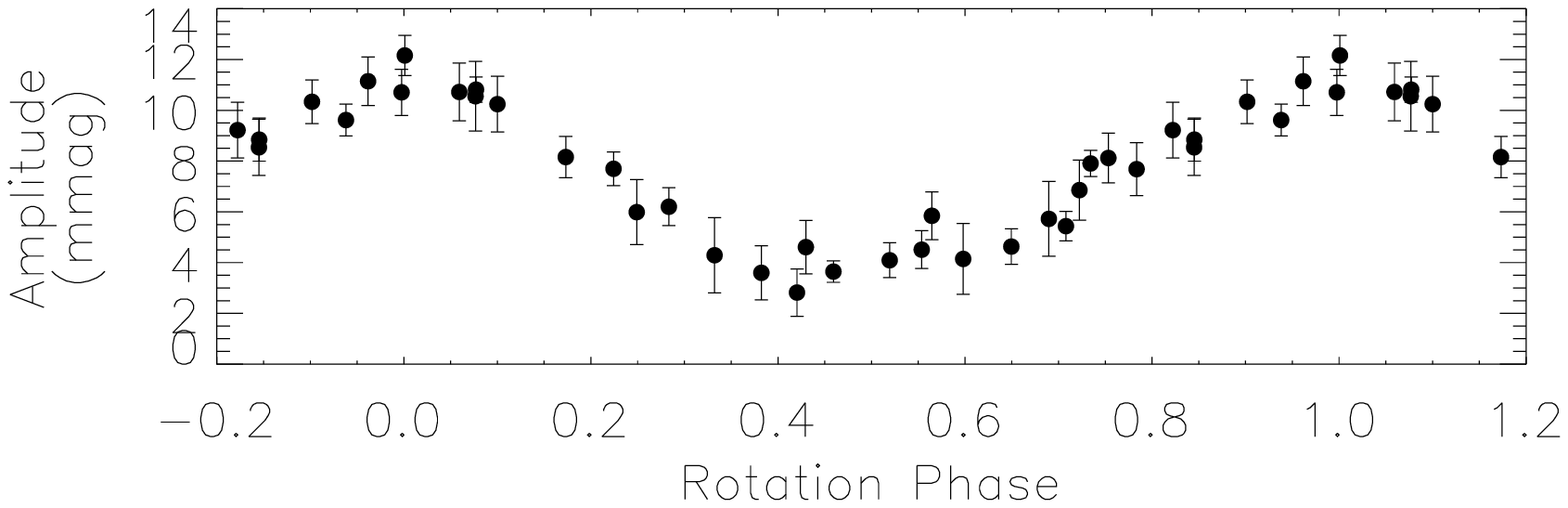}
\includegraphics[width=\linewidth]{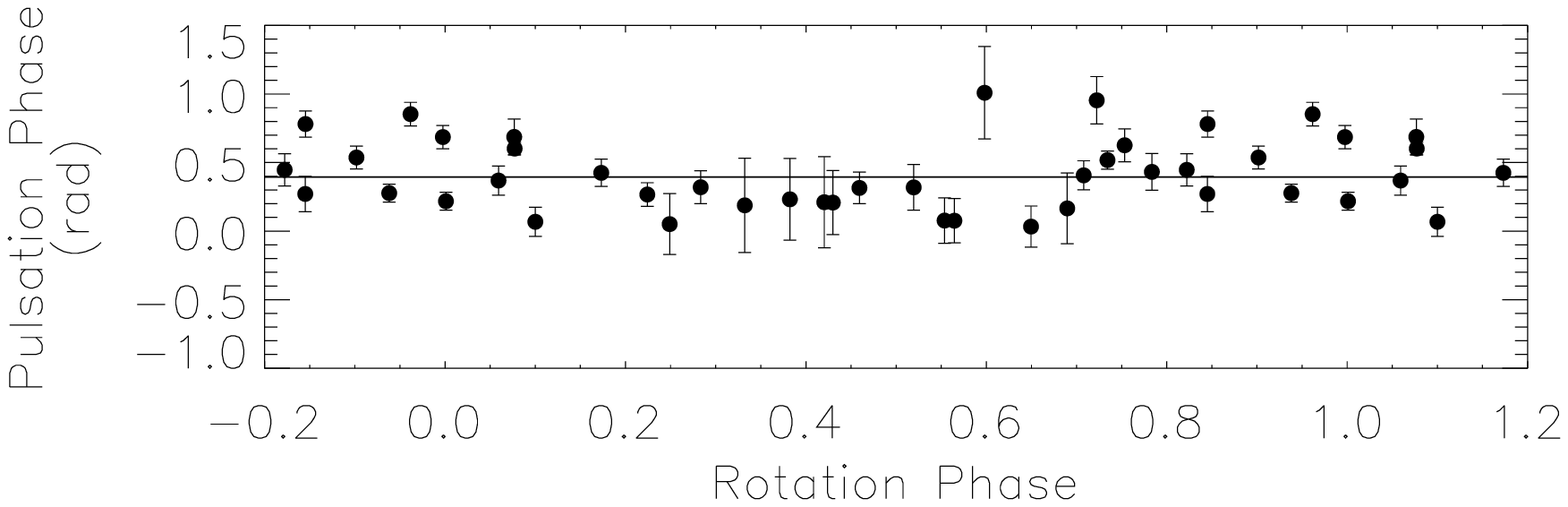}
\caption{Results for testing amplitude and phase variations in the pulsation mode. Top: amplitude variations over the rotation cycle of the star. Bottom: the corresponding phase variations of the pulsation. The amplitude variations are as expected, while the near constant phase suggests a node never crosses the line of sight. The horizontal line in the bottom panel is to guide the eye only. The rotation phase has been calculated such that maximum pulsation amplitude occurs at $\phi_{\rm rot}=0$.}
\label{fig:phamp}
\end{figure}

To perform this test, we take each block of data and calculate the amplitude and phase of the pulsation at a fixed frequency. The results of this test are shown in Fig.\,\ref{fig:phamp}. The observations were constructed such that each visit to the target would yield about 20 pulsation cycles. In some instances, where observations were aborted due to adverse weather or over ridden by higher priority proposals, we have fewer pulsation cycles to fit. This is evidenced by the larger error bars in Fig.\,\ref{fig:phamp}.

It is clear that the amplitude is strongly modulated over the rotation period, ranging from a semi-amplitude of 12\,mmag to about 3\,mmag. The fact that the amplitude of the pulsation never goes to zero confirms our earlier conclusion that $i+\beta\lesssim90^{\circ}$ so that, in the case of a pure dipole mode, the equatorial node never crosses the line of sight. Equally, for a pure quadrupole mode, and a non-zero minimum amplitude, $i+\beta$ must be less than $54.7^{\circ}$, the co-latitude of quadrupole nodes. Furthermore, the phase of the pulsation is almost stable over the rotation period. If $i+\beta > 90^\circ$ (or $54.7^\circ$ for a quadrupole mode) then we would expect a pulsational phase change of $\pi$-rad at quadrature, which is not present in Fig.\,\ref{fig:phamp}.

\section{Pulsation modelling}
\label{sec:modelling}

\begin{figure*}
\includegraphics[width=0.49\linewidth]{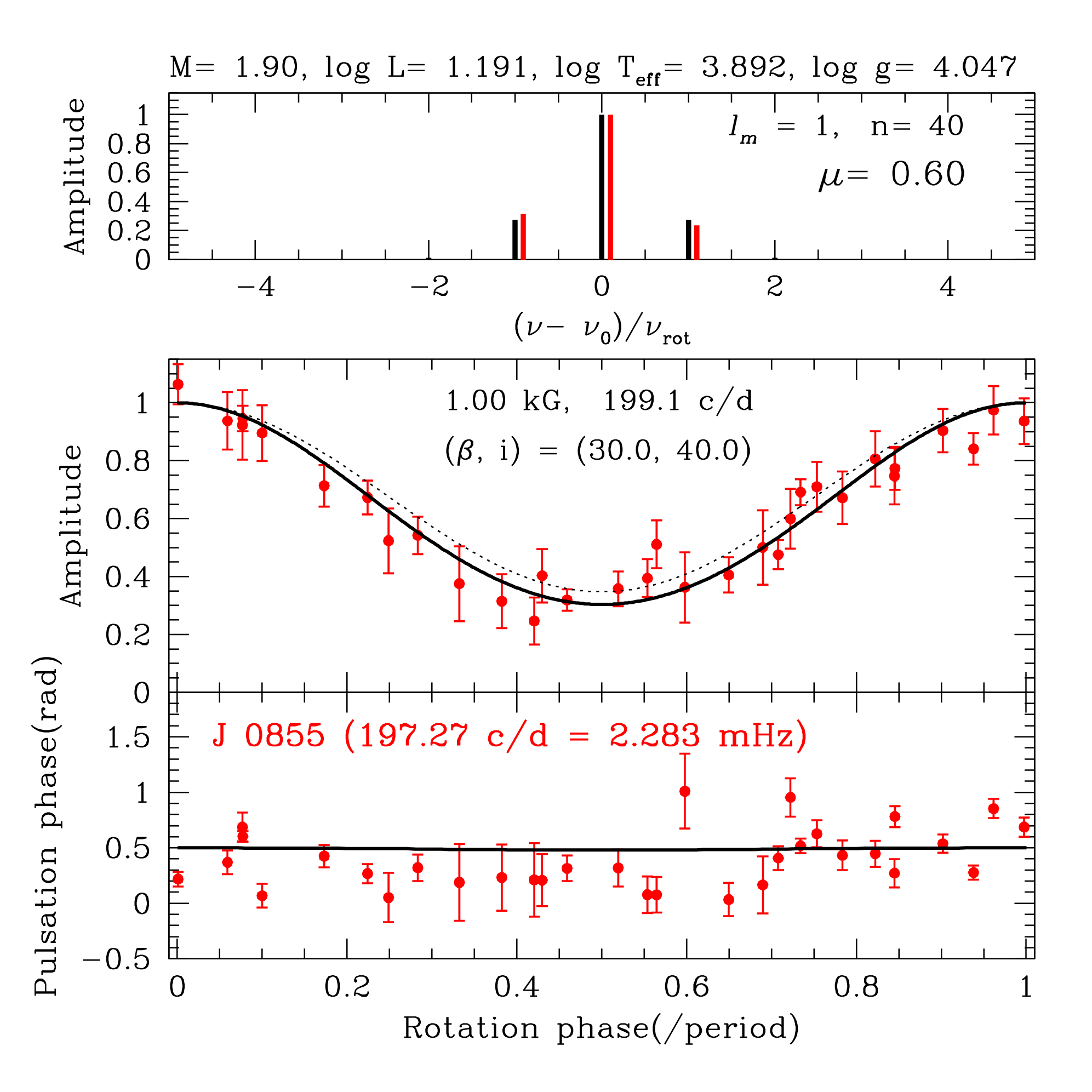}
\includegraphics[width=0.49\linewidth]{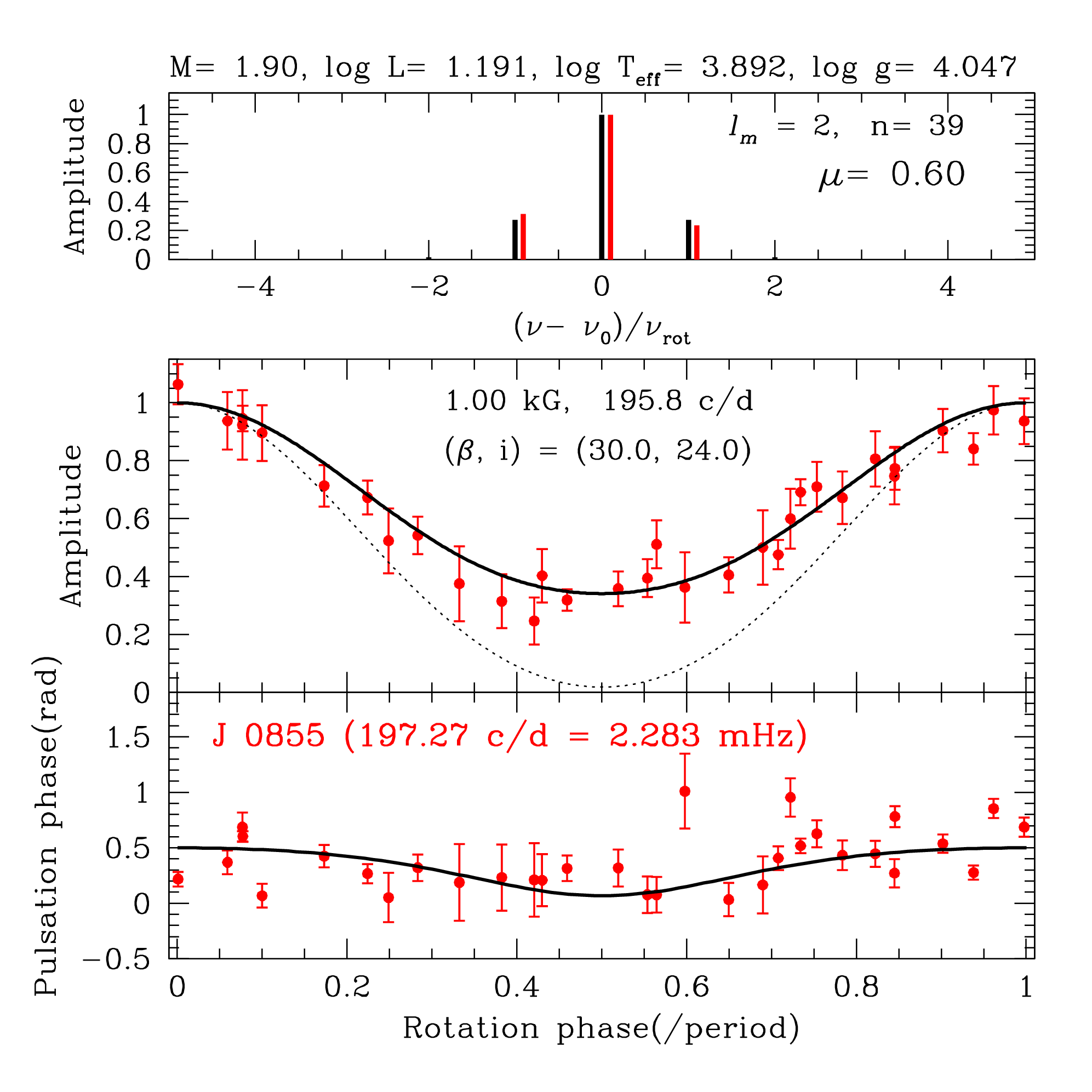}
\caption{The best fitting models for J0855. On the left we present the dipole model, with the quadrupole model on the right. In each case, the red points represent the observations, and the black lines represent the model fit. The dashed lines in the amplitude fit represent the fit of a non-distorted mode. The values of $i$ and $\beta$ are interchangeable without causing a change to the fit.}
\label{fig:model}
\end{figure*}

To test our conclusions derived from the observed data, i.e. the dipole nature of the pulsation and the $i$ and $\beta$ relations, we model the amplitude and phase variations following the recipe of \citet{saio05} and previously employed by \citet{holdsworth16,holdsworth18a}. We refer the interested reader to those papers for full details, but provide a short summary here for completeness. Assuming a dipole magnetic field and stellar limb darkening for an Eddington grey atmosphere (i.e. $\mu=0.6$), we solve the eigenvalue problem for non-adiabatic linear pulsations. For the purpose of the modelling, we take the frequencies and amplitudes from Table\,\ref{tab:freqs} and the effective temperature derived by \citet{holdsworth14a} of $T_{\rm eff}=7800\pm200$\,K. For a selection of masses, 1.8, 1.9, 2.0 and 2.1\,\Msun, an evolutionary model is calculated and a pulsation mode with a frequency close to $\nu$ is found, considering the effects of the magnetic field. A variety of models are calculated, with the constraint that the theoretical amplitude ratio, $0.5[{A_{-1}+A_{+1}}] /{A_0}$ recreates the observed ratio. We present two models in Fig.\,\ref{fig:model} which represent the observations well.

The models for a distorted dipole or quadrupole mode both represent the amplitude modulation well. The phase variations are better represented by the distorted quadrupole mode. However, as the data are relatively sparse we are unsure whether the variation we see is real, or a statistical fluctuation. If it is a statistical fluctuation, then either model is consistent with the observations.

On the other hand, as high order p-mode pulsations are nearly radial, the pulsations of high-order dipole modes and distorted  quadruple modes should be similar except for symmetry of the pulsation phase with respect to the equator. A quadrupole mode distorted by the presence of a spherical symmetric component should be, in general, more visible than a dipole mode, if the two modes are excited to the same amplitude. This leads us to favour the quadrupole mode, but with reservation.

In both cases, we find the following parameters of J0855: $M=1.9$\,\Msun, $\log L=1.19$\,\Lsun, and $B_p$, the polar magnetic field strength, of $1.0$\,kG. Although the theoretical pulsation property depends only weakly on the stellar mass, we chose a $1.9$\,\Msun\, model for J0855 as shown in Fig.\,\ref{fig:model}. This is because the {\it Gaia} DR2 parallax \citep{gaia16,gaiaDR2} gives $\log L=1.16\pm0.04$\,\Lsun, which is similar to the $1.9$\,\Msun\, model at the adopted $T_{\rm eff}$  of $\simeq7800$\,K.
The values of $i$ and $\beta$, which are interchangeable, differ between the models, and are shown on the plots in Fig.\,\ref{fig:model}. The models also predict the expected amplitude of the $\nu\pm2\nu_{\rm rot}$ side lobes to be approximately 0.1\,mmag.

\section{Discussion}

The modelling of J0855 provides us with stellar parameters which allow us to place this star in the context of other roAp stars. Given the mass and luminosity derived from the models, we find that J0855 is pulsating with a frequency much greater than its theoretical acoustic cutoff frequency of about 155\,\cd. The problem of observed pulsations in roAp stars being greater than their cutoff frequencies has been discussed at length in the literature \citep[e.g.][]{ss85a,audard98,gautschy98,sousa08,sousa11}. It has been suggested that the temperature-optical depth relation used to model the Ap stars is not adequate, however modifications of this relationship have not been able to fully explain the phenomenon. We still do not know how these super-critical oscillations are excited in the roAp stars.

\begin{figure}
\includegraphics[width=\linewidth]{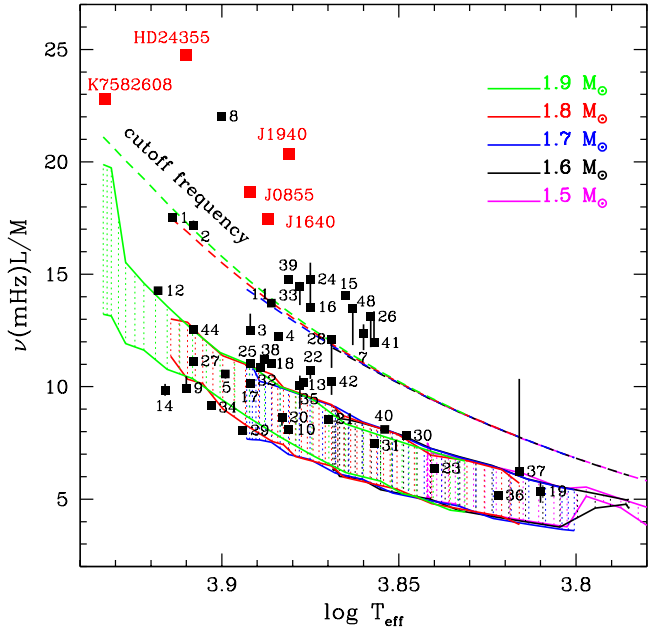}
\caption{The $T_{\rm eff} - \nu L/M$ plane showing the positions of the roAp stars. The plot is based on that of \citet{saio14} and updated from \citet{holdsworth18b} using {\textit{Gaia}} DR2 parallaxes. The squares represent the principle pulsation frequency, with vertical bars showing the range of frequencies for multi-periodic stars. The hatched region represents where high-order p-modes are excited by the $\kappa$-mechanism in the H-ionization zone in non-magnetic models. Acoustic cutoff frequencies are represented by the dashed lines. The numerical labels correspond to the stars in Table\,A1 of \citet{holdsworth18b}.}
\label{fig:teff-LM}
\end{figure}

Further to the realisation of the super-critical pulsation in J0855, we are able to use the model parameters to calculate a $\nu L/M$ value of 18.7 for this star (where $\nu$ is the pulsation frequency in mHz, luminosity L and mass
M are in solar units), placing it close to the distorted quadrupole pulsator J1640 in the $T_{\rm eff} - \nu L/M$ plane, as shown in Fig.\,\ref{fig:teff-LM}. To produce Fig.\,\ref{fig:teff-LM}, we take the values presented in Table\,A1 of \citet{holdsworth18b} and update the luminosities using parallaxes from {\textit{Gaia}} DR2 \citep{gaia16,gaiaDR2} where available and where the error is no more than 5\,per\,cent. For the stars shown in red, we use the best model parameters. We also include a new star in the figure, KIC\,8677585 labeled 48, which a reliable parallax now allows us to plot.

In the region of parameter space surrounding J0855, several other distorted pulsators can be found. All of those stars (plotted in red) are distorted quadrupole pulsators. J0855's proximity to these stars leads us to more heavily support the quadrupole model presented in Fig.\,\ref{fig:model}.

As such, we searched for side lobes at $\nu\pm2\nu_{\rm rot}$ in the hope of identifying a quintuplet. To do this we create a quintuplet with components split by exactly the rotation frequency and fit these by linear least-squares to the light curve. The fit produces no significant detection of signals at these frequencies; we find peaks at 1.0 and 2.3\,$\sigma$ (0.17 and 0.39\,mmag) for the positive and negative $2\nu_{\rm rot}$ components, respectively. The more significant detection of the $-2\nu_{\rm rot}$ is most likely a result of the Coriolis force, if the detection is real. Further observations are needed to reduce the noise and make a significant detection.

All of the distorted pulsators identified in the the $T_{\rm eff} - \nu L/M$ plane have high amplitudes, now including J0855, and strongly suppressed pulsation phase variations. We are still unsure as to whether these two parameters are linked, or whether such a high amplitude has allowed us to identify the peculiarities in these stars and as such we are currently suffering from a selection bias. Either way, an increased number of these high-amplitude distorted quadrupole pulsators will provide a sufficient sample to investigate these stars with a homogenous data set as will be provided by the {\it{TESS}} mission.

{\it{TESS}} observations may also detect the presence of harmonics of the pulsation. Many roAp stars show harmonics of their principal pulsation mode frequency (e.g. $\gamma$\,Equ, \citealt{gruberbauer08}; HD\,99563, \citealt{handler06}; HR\,3831, \citealt{kurtz92}) which demonstrates the non-linearity of their pulsations. It is interesting that, even at the level of precision we have obtained here, we do not see harmonics of the pulsation in J0855. The pulsation amplitude is similar to that seen in J1640 where harmonics up to $3\nu$ were detected \citep{holdsworth18b}, albeit only one had an amplitude above our noise level here. It is still unclear what information can be extracted from the harmonics of the oscillations in some roAp stars, and indeed why some stars show them whilst others do not. Further investigation into this phenomenon may prove insightful in the study of the roAp stars.

\section{Summary and Conclusions}

We have presented 60 hrs of new observations of the rapidly oscillating Ap star J0855 obtained with the Las Cumbres Observatory's 0.4-m telescopes. We have derived a more precise value for the rotation period of this star than was presented in the discovery paper; we determine the period to be $3.0892\pm0.0017$\,d. There is no strong evidence that we see both of the magnetic poles in this star, suggesting that $i+\beta<90^\circ$, with theoretical modelling supporting this conclusion.

Our $B$ observations have revealed this star to be pulsating with a maximum peak-to-peak amplitude of 24\,mmag, placing it among the highest amplitude stars of its class. The pulsation mode is split into a clear triplet with a separation equal to the rotation frequency of the star, as predicted by the oblique pulsator model. There is, however, a tentative detection of a further side lobe to the pulsation at a spacing of $2\nu_{\rm rot}$ which should be confirmed with further observations. We have analysed the pulsation amplitude and phase over the rotation period of the star and found significant phase suppression, as is seen in some other roAp stars.

To understand the phase variations, we have modelled the star and found that a 1.0\,kG polar magnetic field strength is sufficient to distort the mode (given $M=1.9$\,\Msun, $\log L=1.91$\,\Lsun, $\log T_{\rm eff}=3.89$\,K). With the current observations, both a dipole or quadrupole model recreates the variations sufficiently, though we argue in favour of the quadrupole mode on the grounds of mode visibility and the similarity of J0855 to other distorted quadrupole pulsators, especially when considering its position in the $T_{\rm eff}-\nu L/M$ plane. 

\section*{Acknowledgements}

DLH acknowledges financial support from the STFC via grant ST/M000877/1 and the National Research Foundation of South Africa. 
The research leading to these results has received funding from the European Research Council (ERC) under the European Union's Horizon 2020 research and innovation programme (grant agreement N$^{\rm o}$ 670519: MAMSIE). 
This paper uses observations made with the Las Cumbres Observatory (LCO) 0.4-m telescope network. 
This work has made use of data from the European Space Agency (ESA) mission
{\it Gaia} (\url{https://www.cosmos.esa.int/gaia}), processed by the {\it Gaia}
Data Processing and Analysis Consortium (DPAC,
\url{https://www.cosmos.esa.int/web/gaia/dpac/consortium}). Funding for the DPAC
has been provided by national institutions, in particular the institutions
participating in the {\it Gaia} Multilateral Agreement.
We thank Prof. Donald W. Kurtz for constructive comments on the manuscript.
We thank the anonymous referee for useful suggestions to the manuscript.

\bibliography{J0855-refs}

\label{lastpage}
\end{document}